\documentclass[twocolumn,aps,prd,showpacs,superscriptaddress,longbibliography]{revtex4-1}

\usepackage{amsmath}
\usepackage{amssymb}
\usepackage{cancel}
\usepackage{maybemath}
\usepackage{xcolor}
\usepackage{graphicx}
\usepackage{bm}

\usepackage[%
  colorlinks=true,
  urlcolor=blue,
  linkcolor=blue,
  citecolor=blue
]{hyperref}
  
\usepackage{dcolumn}
\newcolumntype{X}{D{.}{.}{23}}

\newcommand{\dd}{\mathrm{d}}
\newcommand{\ii}{\mathrm{i}}

\newcommand{\calM}{\mathcal{M}}

\newcommand{\calT}{\mathcal{T}}

\newcommand{\thr}{\mathrm{th}}

\begin{document}

\title{Lepton-pair \v{C}erenkov radiation emitted by tachyonic neutrinos:\\
Lorentz-covariant approach and IceCube data}

\author{Ulrich D. Jentschura}

\affiliation{Department of Physics, Missouri University of Science and
Technology, Rolla, Missouri 65409, USA}

\author{Robert Ehrlich}

\affiliation{Department of Physics, George Mason University, Fairfax, Virginia
22030, USA}

\begin{abstract}
{\bf Abstract:}
Current experiments do not exclude the possibility that one or more neutrinos
are very slightly superluminal or that they have a very small tachyonic mass.
Important bounds on the size of a hypothetical tachyonic neutrino
mass term are set by lepton pair \v{C}erenkov radiation (LPCR), i.e.,
by the decay channel $\nu \rightarrow e^+e^-\nu$ which
proceeds via a virtual $Z^0$ boson.
Here, we use a Lorentz-invariant dispersion relation which leads to very tight
constraints on the tachyonic mass of neutrinos;
we also calculate decay and energy loss rates.
A possible cutoff seen in the IceCube neutrino
spectrum for $E_\nu >2 \, {\rm PeV}$, due to the
potential onset of LPCR, is discussed. 
\end{abstract}


\maketitle

%
% Introduction
%
\section{Introduction}
\label{sec1}

The early arrival of a neutrino burst from the 1987A supernova~\cite{DaEtAl1987}
still motivates speculations about a possible superluminal nature 
of neutrinos, even if it is generally assumed that the 
delay in the arrival of electromagnetic radiation (light) 
is caused by the time the shock wave from the core collapse needs in
order to reach the surface of the exploding star.
If neutrinos are ever so slightly superluminal, then 
they may emit \v{C}erenkov radiation in the form of 
light lepton pairs.
In this paper, we attempt to answer three questions:
{\em (i).}~How would the energy threshold for the decay channel
$\nu \rightarrow e^+e^-\nu$ (lepton pair \v{C}erenkov radiation, LPCR)
have to be calculated if we assume a strictly Lorentz-covariant,
space-like dispersion relation for the relevant neutrino flavor
eigenstate?
{\em (ii).}~How would the decay rate and the energy loss rate have
to be calculated under this assumption?
Can the tachyonic Dirac equation~\cite{ChHaKo1985,Ch2000,Ch2002,Ch2002a} 
and its bispinor solutions~\cite{JeWu2012epjc,JeWu2013isrn}
be used in that context?
{\em (iii).}~What implications could be derived for astrophysics, under the
assumption that a possible cutoff seen by IceCube 
for neutrinos with energies $E_\nu>2 \, {\rm PeV}$ is confirmed 
by future experiments?

Theoretical arguments can be useful in restricting the possible degree of
superluminality of neutrinos and maximum attainable neutrino
velocities~\cite{CoGl2011,BeLe2012,DaMo2012}. In
Refs.~\cite{CoGl2011,BeLe2012}, a Lorentz-noncovariant 
dispersion relation $E_\nu = |\vec p|\, v_\nu$ was used,
where $v_\nu > c$ is a constant parameter.
This assumption leads to an energy-dependent effective ``mass''
square $E_\nu^2 - \vec p^{\,2} \approx E^2_\nu \, (v_\nu^2 - 1) \, v_\nu^{-2}
\equiv m_{\rm eff}^2$.
The effective mass $m_{\rm eff} = E_\nu \, \sqrt{v_\nu^2-1} \, v_\nu^{-1}$ 
then grows linearly with the neutrino energy.
(Natural units with $\hbar = c = \epsilon_0 = 1$ are used in this paper,
yet we shall include explicit factors of $c$ when indicated
by the context.)
Indeed, at the time, a best fit to the available experimental
neutrino mass data including the initial OPERA claim~\cite{OPERA2011v1}
suggested the conceivable existence of 
an ``energy-dependent mass'' of the neutrino,
as evidenced in Fig.~1 of Ref.~\cite{TaLa2012}.
The choice of the relation $E_\nu = |\vec p|\, v_\nu$ 
made in Ref.~\cite{CoGl2011}
was consistent with the need to model the initial OPERA claim~\cite{OPERA2011v1},
and is perfectly compatible with the concept of
perturbative Lorentz breaking terms in the neutrino sector~\cite{BeLe2012}.
A Dirac-type equation leading to the Lorentz-noncovariant 
dispersion relation used by Cohen and Glashow~\cite{CoGl2011}
can be obtained~\cite{BeLe2012}
from the current operator given in Eq.~(2) of Ref.~\cite{KoLe2001}
upon a particular choice of the $c^{\mu\nu}$ parameters 
in the generalized fermionic current operator
(in the notation adopted in Ref.~\cite{KoLe2001}).
Then, assuming a constant neutrino speed $v_\nu > c$, 
one can effectively describe the apparent absence of an energy dependence of the
deviation of the neutrino speed from the speed of light $v_\nu \approx
\mathrm{const.} \gtrsim c$ (in the range $5 \, {\rm GeV} < E_\nu < 50 \, {\rm
GeV}$), according to the (falsified)
initial claim made by OPERA~\cite{OPERA2011v1},
while remaining comptabile with the framework of 
perturbative Lorentz breaking~\cite{KoLe2001}.

However, while there are advantages to assuming 
a Lorentz-noninvariant dispersion relation for 
superluminal neutrinos (such as the preservation of the 
timelike positive quantity $E_\nu^2 - \vec p^{\,2} > 0$),
there are also a number of disadvantages.
For example, if the dispersion relation $E_\nu = |\vec p|\, v_\nu$ 
holds in one particular Lorentz frame, then under a Lorentz boost, in general,
one has $E'_\nu \neq |\vec p'|\, v_\nu$  in the moving frame~\cite{CoGl2011,BeLe2012}. 
In order to illustrate the consequences 
of Lorentz noncovariance, let us consider a boost along the positive $z$
axis into a frame which moves with
velocity $u = c^2/v_\nu < c$.
A particle moving along the positive $z$ axis of the lab frame
with four-momentum $p^\mu = (|\vec
p| v_\nu, |\vec p| \,\hat e_z)$ is mapped onto $p'^\mu = (|\vec p| \,
\sqrt{v_\nu^2 - 1}, \vec 0)$ and thus is ``at rest'' in the moving frame.
However, the general dispersion relation in the moving frame,
\begin{equation}
\label{eprime}
E'_\nu = -\frac{p'_z}{2 v_\nu} -
\frac{(p'^2_x + p'^2_y + p'^2_z) \, v_\nu}{2 p'_z}
\qquad (p'_z \neq 0)\,,
\end{equation}
is much more complicated.
(Throughout this paper, we denote the spatial components of the
four-vector $p^\mu = (E_\nu, \vec p)$ by $\vec p$ and keep $|\vec p|$ explicitly,
in order to avoid confusion between $p^2 = p^\mu p_\mu$ with $p^2 \neq \vec
p^{\,2}$.)

An alternative, commonly accepted dispersion relation for 
so-called tachyons (these are space-like,
faster-than-light particles
described by a Lorentz-invariant wave equation) reads as $E_\nu^2 = \vec p^{\,2} -
m_\nu^2$, i.e., it is the ``normal'' dispersion relation with the negative sign
of the mass square term (see
Refs.~\cite{BiDeSu1962,DhSu1968,BiSu1969,Fe1967,Fe1978,%
ChHaKo1985,Ch2000,Ch2002,Ch2002a,JeWu2012epjc,JeWu2013isrn,JeWu2014,%
Eh1999a,Eh1999b,Eh2012,Eh2013}).  
Here, we calculate the
threshold energy and the decay rate under the assumption
of a Lorentz-invariant dispersion relation for the neutrino.  
We find that the alternate
dispersion relation imposes a tight restrictions on superluminality,
and has important phenomenological implications for neutrino masses.

%
% Dispersion Relations
%
\section{Dispersion Relations and Thresholds}
\label{sec2}

For tachyonic particles, starting from the pioneering work 
of Sudarshan {\em et al.}~\cite{BiDeSu1962,DhSu1968,BiSu1969},
continuing with the works of Feinberg~\cite{Fe1967,Fe1978},
and including the tachyonic neutrino 
hypothesis~\cite{ChHaKo1985,Ch2000,Ch2002,Ch2002a,%
JeWu2012epjc,JeWu2014,Eh1999a,Eh1999b,Eh2012,Eh2013},
the following dispersion relation has been assumed for the 
tachyonic (space-like) solutions,
\begin{subequations}
\begin{align}
\label{spencer1}
E_\nu =& \; \gamma_\nu \, m_\nu \,,
\qquad
| \vec p_\nu | = \gamma_\nu \, m_\nu \, v_\nu \,,
\\[0.133ex]
\label{spencer2}
|\vec p_\nu| =& \; E_\nu \, v_\nu \,,
\qquad
p^\mu \, p_\mu = E_\nu^2  - \vec p_\nu^{\,2} = -m_\nu^2  \,,
\end{align}
\end{subequations}
where we use the suggestive subscript $\nu$ for ``neutrino''.
These relations imply that $| \vec p| = E_\nu \, v_\nu$
instead of $E_\nu = |\vec p| \, v_\nu$.
Here, the tachyonic Lorentz factor appears which is 
$\gamma_\nu = 1/\sqrt{v_\nu^2 - 1}$.
Tachyonic and tardyonic dispersion relations are unified 
upon assuming an imaginary value for $m$ in the tachyonic case
(starting from the tardyonic 
case, one has $E = m/\sqrt{1-v^2} \to \ii m/\sqrt{1-v^2} = m/\sqrt{v^2-1}$,
where the latter equation holds for tachyons). 
With the standard definitions of $\vec p$ and $E_\nu$, one has
$|\vec p_\nu| = \gamma_\nu m \, v_\nu = E_\nu \, v_\nu$
for both tardyons and tachyons. 

\begin{figure}[t!]
\begin{center}
\begin{minipage}{0.99\linewidth}
\begin{center}
\includegraphics[width=0.6\linewidth]{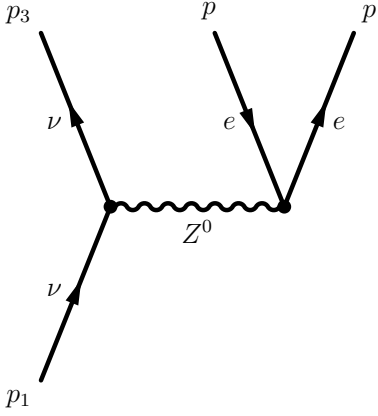}
\caption{\label{fig1}
Conventions for tachyonic neutrino decay.}
\end{center}
\end{minipage}
\end{center}
\end{figure}

In order to obtain the threshold energy for the 
LPCR decay $\nu\rightarrow e^+e^-\nu$, 
we use the following conventions (see Fig.~\ref{fig1}),
inspired by Chap.~10 of Ref.~\cite{Gr1987}, 
and define $E_1= \sqrt{\vec p_1^{\,2} - m_\nu^2}$ and $E_3=
\sqrt{\vec p_3^{\,2} - m_\nu^2}$ as the oncoming and 
outgoing neutrino energies, with 
$q=(E_1, \vec p_1)-(E_3, \vec p_3)$ being the
four-momentum of the $Z^0$.  Pair production threshold is reached for 
$q^2 = 4m_e^2$ and 
$\cos \theta= \vec{p_1} \cdot \vec{p_3}/(|\vec p_1| \, |\vec p_3| ) =1$.
For collinear geometry,
with all momenta pointing along the $z$ axis, we have
\begin{align}
\label{threshold}
q^2 =& \; \left(\sqrt{p_{1z}^2 - m_\nu^2} - \sqrt{p_{3z}^2 - m_\nu^2}\right)^2 
\nonumber\\[0.133ex]
& \; - ( p_{1z} - p_{3z} )^2  = 4m_e^2 \,.
\end{align}
Furthermore, threshold obviously requires $E_3=0$.
(This is possible for tachyonic particles, when $|\vec p_3| = p_{3z} = m_\nu$.
In this limit, the tachyonic particles becomes infinitely 
fast, and loses all of its energy, which implies that it is 
impossible to detect it~\cite{OULU_WEB2}. The counterintuitive loss of 
energy for tachyons under acceleration is a consequence 
of standard tachyonic kinematics~\cite{BiDeSu1962,DhSu1968,BiSu1969,%
Fe1967,Fe1978,ChHaKo1985,Re2009,Bi2009,Bo2009,JeWu2012epjc,JeWu2013isrn}.)
When the relations $E_3=0$ and $|\vec p_3| = p_{3z} = m_\nu$ 
are substituted into Eq.~\eqref{threshold}, this yields
\begin{equation}
p_{1z}^2 - m_\nu^2 - ( p_{1z} - m_\nu )^2  = 4m_e^2 \,.
\end{equation}
Identifying $p_{1z} = | \vec p |_\thr$ with the threshold momentum,
one easily finds
\begin{equation}
| \vec p |_\thr = \frac{2 m_e^2}{m_\nu} + m_\nu \,.
\end{equation}
The threshold energy is then easily found as
\begin{equation}
\label{Ethr}
E_{\thr} =
\sqrt{ \vec p^{\,2}_\thr - m_\nu^2 } =
2 \, \frac{m_e}{m_\nu}  \, \sqrt{ m_e^2 + m_\nu^2}
\approx
2 \, \frac{m_e^2}{m_\nu} \,.
\end{equation}
Because we are using a tachyonic dispersion relation,
the threshold energy can be expressed a function of only the mass parameters.
Larger tachyonic masses $m_\nu$ lead to lower threshold energies.
In view of the tachyonic dispersion relation 
$m_\nu = E_\thr \, \sqrt{ v_\thr^2 - 1 }$, where $v_\thr$ is the 
neutrino velocity {\em at threshold}, we may convert the 
threshold energy into a function of the electron mass and the neutrino 
threshold velocity.
For given $E_\nu$, the limit $m_\nu \ll m_e$ is equivalent to the 
limit $ v_\thr^2 - 1 = \delta_\thr \to 0$ 
because $m_\nu = E_\nu \, \sqrt{ \delta_\thr }$.\
In this limit, we have
\begin{align}
\label{mainresapprox}
E_{\thr} \approx & \;
2 \, \frac{m_e^2}{m_\nu} =
2 \, \frac{m_e^2}{E_\thr \, \sqrt{ v_\thr^2 - 1 }}
\nonumber\\[2ex]
\Rightarrow & \;
E_\thr \approx
\frac{\sqrt{2} \, m_e}{( v_\thr^2 - 1 )^{1/4}} \,.
\end{align}
Substituting the exact dispersion relation 
into the threshold condition 
$E_{\thr} = 2 \, \frac{m_e}{m_\nu}  \, \sqrt{ m_e^2 + m_\nu^2}$,
and solving for $E_\thr$, one obtains
\begin{align}
\label{mainres}
E_\thr =& \; \sqrt{2} \, m_e \, \left( 1 + 
\frac{v_\thr}{\sqrt{v_\thr^2 - 1}} \right)^{1/2}
\nonumber\\[0.133ex]
= & \;
\left\{ \begin{array}{cc}
\dfrac{\sqrt{2}\, m_e}{\delta_\thr^{1/4}} \qquad & \delta_\thr \ll 1 
\\
2 \,  m_e + \dfrac{m_e}{4 \, \delta_\thr} \qquad & \delta_\thr \gg 1
\end{array}
\right. \,.
\end{align} 
The exact expression~\eqref{mainres} confirms Eq.~\eqref{mainresapprox} 
in the limit $\delta_\nu \ll 1$, which corresponds to 
the phenomenologically important limit of 
high-energy neutrinos. Smaller values of $\delta_\thr$ 
(approaching zero) correspond to smaller 
tachyonic neutrino masses and therefore, to 
larger threshold energies. 
For given neutrino speed $v_\thr$, 
neutrinos with energy 
$E_\thr$ (or larger), under the hypothetical assumption of
the tachyonic dispersion relation, have a tachyonic 
neutrino mass term large enough to make the decay via LPCR
kinematically possible.
Expressed differently, the tachyonic mass term
$m_\nu = E_\thr \, \sqrt{ v_\thr^2 - 1 }$ in this case
is large enough to lead to LPCR decay at energy $E_\thr$,
according to Eq.~\eqref{Ethr}.

\begin{figure}[t!]
\begin{center}
\begin{minipage}{0.99\linewidth}
\begin{center}
\includegraphics[width=0.8\linewidth]{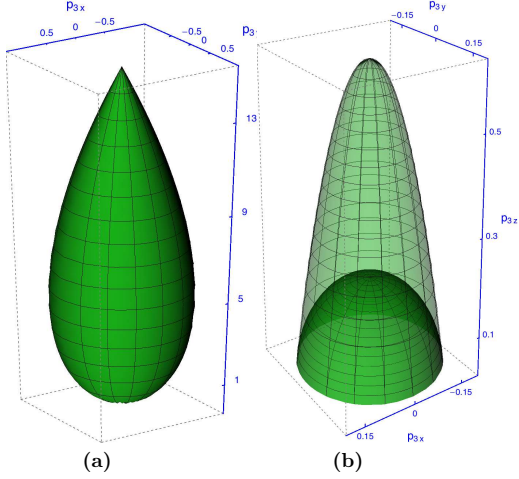}
\caption{\label{fig2} 
Fig.~(a): Region of allowed outgoing momenta $\vec p_3$
for the decay of an incoming superluminal neutrino with 
$E_\nu = |\vec p| \, v_\nu$.
The neutrino is incoming
along the positive $z$ axis ($p_{1z} = 15$).
The boundary of allowed $\vec p_3$ vectors constitutes
a distorted ellipsoid with a ``sharpened tip'',
obtained as a solution of setting $q^2 = 0$
in Eq.~\eqref{q2super}.
Fig.~(b): Region of allowed
$\vec p_3$ vectors for an incoming tachyonic neutrino with $p_{1z} = 62$
and $-m_\nu^2 = -(0.2)^2$, producing an electron-positron
pair of mass $m_e = 1$ (dispersion relation 
$E_\nu = \sqrt{ \vec p_\nu^{\,2} - m_\nu^2}$).
Final wave vectors $|\vec p_3| < m_\nu$ correspond to evanescent waves and are
thus to be excluded~\cite{JeWu2012epjc}.}
\end{center}
\end{minipage}
\end{center}
\end{figure}

%
% Decay Rate and Time--Like Dispersion Relation
%
\section{Decay Rate and Time--Like Noncovariant Dispersion Relation}
\label{sec3}

Given the complexities of calculating the decay rate
due to LPCR using a tachyonic dispersion
relation, it is extremely useful to 
first discuss the case of a Lorentz noncovariant 
form $E_\nu = |\vec p| \, v_\nu$,
using lab frame variables.
For collinear incoming and outgoing neutrinos,
threshold for pair production is reached at
$q^2 = ( E_1 - E_3)^2 - (p_{1z} - p_{3z})^2 
=(p_{1z} - p_{3z})^2 \, (v_\nu^2 - 1) = 4 m_e^2$,
from which one derives (setting $\vec p_3 = \vec 0$) the 
following threshold values (in agreement with Ref.~\cite{CoGl2011}),
\begin{equation}
| \vec p_1 |_{\thr} = \frac{ 2 m_e }{\sqrt{ v_\nu^2 - 1 }} \,,
\qquad
(E_1)_{\thr} = \frac{ 2 m_e \, v_\nu }{\sqrt{ v_\nu^2 - 1 }} \,.
\end{equation}
Here, $G_F$ is Fermi's coupling constant
and the $u$ and $v$ are the standard fundamental
positive-energy and negative-energy bispinor solutions
of the Dirac equation~\cite{ItZu1980}.
The invariant matrix element is 
\begin{align}
\calM =& \; \frac{G_F}{\sqrt{2}} \,
\left[ \overline u(p_3) \, \gamma_\lambda \, ( 1- \gamma^5) \,
u(p_1) \right] \,
\nonumber\\[0.1133ex]
& \; \times \left[ \overline u(p_4) \, \left( c_V \gamma_\lambda -
c_A \, \gamma_\lambda \, \gamma^5 \right) \, v(p_2) \right] \,.
\end{align}
Here, $c_V \approx 0$, and $c_A \approx -1/2$
[see Eq.~(5.57) on p.~153 of Ref.~\cite{Ho2002}].
Following~\cite{BeLe2012}, we now make the 
additional assumption that the 
functional form of the projector sum over the spin orientations
remains the same as for the ordinary Dirac equation
even if the underlying dispersion relation is Lorentz-noncovariant
(for a general discussion on such models see Ref.~\cite{RuSaSi2012,KoMe2012}).
In this case, 
the sum over final state and the averaging over the initial spins leads to
$\frac12 \, \sum_{\rm spins} | \calM |^2 =
64 \, G_F^2 \, (p_1 \cdot p_2) \, (p_3 \cdot p_4)$.
This enters the lab-frame expression for the decay rate~\cite{Jo2011},
\begin{align}
\label{GammaBasic}
\Gamma =& \; \frac{1}{2 E_1} \,
\int \frac{\dd^3 p_3}{(2 \pi)^3 \, 2 E_3}
\, \left(
\int \frac{\dd^3 p_2}{(2 \pi)^3 \, 2 E_2}
\int \frac{\dd^3 p_4}{(2 \pi)^3 \, 2 E_4}
\right.
\nonumber\\[0.1133ex]
& \; \times \left. (2 \pi)^4 \, \delta^{(4)}( p_1 - p_3 - p_2 - p_4 ) \,
\left[ \frac12 \, \sum_{\rm spins} | \calM |^2 \right] \right)
\nonumber\\[0.1133ex]
=& \; \frac{G_F^2}{12 \, \pi^4 \, (2 E_1)} \, \int \frac{\dd^3 p_3}{2 E_3} \,
\left( p_1 \cdot p_3  \, q^2 + 2 \, ( p_1 \cdot q ) \, (p_3 \cdot q) \right) \,,
\end{align}
where $q = p_1-p_3$.
The azimuthal symmetry suggests the 
use of cylindrical coordinates.
The domain of integration contains, for given
$p_1 = (p_{1z} \, v_\nu, 0, 0, p_{1z})$, all permissible
$\vec p_3 = p_{3\rho} \, \hat e_\rho + p_{3z} \hat e_z $,
where $p_3^\mu = (|\vec p_3| v_\nu, \vec p_3)$.
With $E_\nu = |\vec p| \, v_\nu$, 
the momentum transfer is 
\begin{equation}
\label{q2super}
q^2 = -(p_{1z} - p_{3z})^2 - p_{3\rho}^2 +
\left( p_{1z} - 
\sqrt{p_{3\rho}^2 + p_{3z}^2} \right)^2 v_\nu^2 \,,
\end{equation}
where we require $q^2 > 4 m_e^2 \approx 0$.
Solving Eq.~\eqref{q2super} for $p_{3 \rho}$,
one obtains the boundary of the region 
of permissible $\vec p_3$ vectors. 
An example is given in Fig.~\ref{fig2}(a),
in the form of a ``sharpened ellipsoid'' with a ``sharp'' top
near $p_{3\rho} \to 0$, $p_{3z} \to p_{1z}$, and a
``rounded'' bottom with $p_{3\rho} \to 0$,
and $p_{3z} \to - [(v_\nu-1)/(v_\nu+1)] \, p_{1z}$.
After a somewhat tedious integration over the 
allowed $\vec p_3$ vectors, one obtains
\begin{align}
\label{CoGlGammaAnddEdx}
\Gamma =& \;
\frac{G_F^2}{2688 \, \pi^3}  \frac{p_{1z}^5 \delta_\nu^3}{v_\nu}
\approx
\frac{1}{14} \, \frac{G_F^2 \, E_\nu^5 \delta_\nu^3}{192 \, \pi^3} \,,
\nonumber\\[0.1133ex]
\frac{\dd E_\nu}{\dd x} \approx & \;
-\frac{G_F^2}{96 \, \pi^4 \, (2 E_\nu)} \,
\int\limits_{q^2 > 0} \frac{\dd^3 p_3}{2 E_3} \,
(E_\nu - E_3) 
\nonumber\\[0.1133ex]
& \; \times \left[ (p_1 \cdot p_3) \, q^2
+ 2 \, (p_1 \cdot q) (p_3 \cdot q) \right] 
\nonumber\\[0.1133ex]
=& \;
-\frac{G_F^2}{86016 \, \pi^3}  \frac{p_{1z}^6 \delta_\nu^3}{v_\nu}
\approx
-\frac{25}{448} \, \frac{G_F^2 \, E_\nu^6 \, \delta_\nu^3}{192 \, \pi^3} \,,
\end{align}
for the energy loss per unit length,
confirming the results given in Eq.~(2) and~(3)
of~Ref.~\cite{CoGl2011}, and in Ref.~\cite{BeLe2012}.
This confirmation of the results given in Ref.~\cite{CoGl2011}
(under the assumptions made in the cited paper,
namely, the dispersion relation $E_\nu = |\vec p|\, v_\nu$),
but using a different method, namely, phase-space 
integration directly in the laboratory frame,
encourages us to apply the same method to the 
calculation of the tachyonic neutrino decay rate,
where the use of the laboratory frame is indispensable.
The confirmation also underlines the consistency 
of the theoretical formalism under a change of the 
assumptions made in the calculation.

\begin{figure}[t!]
\begin{center}
\begin{minipage}{0.99\linewidth}
\begin{center}
\includegraphics[width=0.7\linewidth]{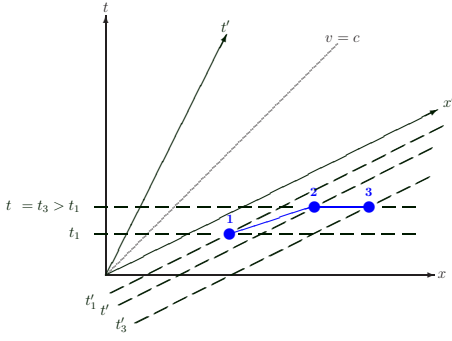}
\end{center}
\caption{\label{fig3} 
The world line $1 \mapsto 2 \mapsto 3$ describes the
tachyonic neutrino decay into a zero-energy,
infinitely fast neutrino.
Complete reversal of the time ordering of the decay
process takes place
in the primed frame; the observer
interprets the process as the decay of an
incoming antineutrino along the trajectory
$3 \mapsto 2 \mapsto 1$.}
\end{minipage}
\end{center}
\end{figure}

%
% Decay Rate and Space--Like Dispersion Relation
%
\section{Decay Rate and Space--Like Covariant Dispersion Relation}
\label{sec4}

For an incoming tachyon, the particle state (space-like neutrino) may
transform into an antiparticle state upon Lorentz transformation, and its
trajectory may reverse the time ordering (see Fig.~\ref{fig3}). 
Thus, the
interpretation of a tachyonic neutrino state as a particle or antiparticle may
depend on the frame of reference, and we should calculate the process directly in the
lab frame.  The necessity to transform certain tachyonic particle field
operators into antiparticle operators under Lorentz boosts has been stressed in
Ref.~\cite{Fe1967,Fe1978,JeWu2012epjc}.  
Incoming and outgoing states are required to be above-threshold
positive-energy states in the lab frame
(causality and tachyonic trajectories are discussed in 
Refs.~\cite{BiDeSu1962,DhSu1968,BiSu1969,%
Fe1967,Fe1978,ChHaKo1985} and Appendix A.2 of Ref.~\cite{JeEtAl2014}). 

We consider the matrix element
\begin{align}
\calM =& \; \frac{G_F}{\sqrt{2}} \,
\left[ \overline u^\calT(p_3) \, \gamma_\lambda \, ( 1- \gamma^5) \,
u^\calT(p_1) \right] 
\nonumber\\[0.1133ex]
& \; \times \left[ \overline u(p_4) \, \left( c_V \gamma^\lambda -
c_A \, \gamma^\lambda \, \gamma^5 \right) \, v(p_2) \right] \,.
\end{align}
Here, the $u^\calT(p_1)$, and $u^\calT(p_3)$ are Dirac
spinor solutions of the tachyonic Dirac 
equation~\cite{JeWu2012epjc,JeWu2013isrn}.
The bar denotes the Dirac adjoint.
In the helicity basis
(see Chap.~23 of Ref.\cite{BeLiPi1982vol4}
and Refs.~\cite{JeWu2012epjc,JeWu2013isrn}), 
these are given by
\begin{equation}
u^\calT_\pm(p) = 
\left( \begin{array}{c} \sqrt{ | \vec p | \pm m} \; a_\pm(\vec p) \\
\pm \sqrt{ | \vec p | \mp m} \; a_\pm(\vec p) 
\end{array} \right) \,,
\end{equation}
where the $a_\pm(\vec p)$ are the fundamental
helicity spinors (see p.~87 of Ref.~\cite{ItZu1980}).
Following~\cite{JeWu2012epjc,JeWu2013isrn,JeWu2014},
we use the tachyonic sum rule of the 
fundamental tachyonic bispinor solutions
[see Eq.~(34a) of Ref.~\cite{JeWu2012epjc}]
\begin{equation}
\label{sumrule1}
\sum_\sigma (-\sigma )\;
u^\calT_\sigma(p) \otimes \overline u^\calT_\sigma(p) \, \gamma^5 =
\cancel{p} - \gamma^5 \, m \,,
\end{equation}
where $p = (E, \vec p)$ is the four-momentum, and
$\sigma$ is a helicity quantum number.
We refer to Refs.~\cite{JeWu2012epjc,JeWu2013isrn}
for a thorough discussion; roughly speaking,
the factor $(-\sigma)$ in Eq.~\eqref{sumrule1}
restores the correct sign in the calculation of the 
time-ordered product of tachyonic field operators
(the propagator), for the contribution of all virtual
degrees of freedom of the tachyonic field
[see Eqs.~(46)--(57) and Eq.~(73)--(75) of 
Ref.~\cite{JeWu2013isrn}].
The $\gamma^5$ matrix in Eq.~\eqref{sumrule1} 
is a part of the natural Dirac ``adjoint'' for the tachyonic 
spinor. Namely, the adjoint equation to the 
tachyonic Dirac equation, 
$(\ii \gamma^\mu \partial_\mu - \gamma^5 m) \psi(x) = 0$,
reads as 
$\left[ \overline \psi(x) \, \gamma^5 \right] \, 
(\ii \gamma^\mu \overleftarrow{\partial}_\mu - \gamma^5 m) \psi(x) = 0$.
As explained in Eqs.~(73)--(75) of
Ref.~\cite{JeWu2013isrn}, right-handed particle 
and left-handed antiparticle states (those with the ``wrong''
helicity) are excluded from the physical spectrum of the
tachyonic field by a Gupta--Bleuler condition;
these cannot contribute to the oncoming and outgoing 
neutrino states in Fig.~\ref{fig1} [while they do 
contribute to the virtual states, i.e., the propagator,
see Eqs.~(46)--(57) of Ref.~\cite{JeWu2013isrn}].
Both the incoming as well as the outgoing neutrinos 
in Fig.~\ref{fig1} are real rather then virtual neutrinos.
Hence, in order to calculate the LPCR decay rate,
we use the modified sum over tachyonic spinors
\begin{align}
\label{sumrule2}
\widetilde\sum_\sigma
u^\calT_\sigma(p) \otimes \overline u^\calT_\sigma(p)
= (1 + \gamma^5 \, \cancel{\tau} \, \cancel{\hat{p}}) \;
( \cancel{p} - \gamma^5 \, m_\nu) \; \gamma^5 \,,
\end{align}
where $\tau = (1,0,0,0)$ is a time-like unit vector,
$\hat p = \vec p / | \vec p|$ is the unit vector in the $\vec p$ direction,
and upon promotion to a four-vector, we have
$\hat p^\mu = (0, \hat p)$, so that 
$1 + \gamma^5 \, \cancel{\tau} \, \cancel{\hat{p}} = 
1 - \vec \Sigma \cdot \vec p/|\vec p|$ becomes
a left-handed helicity projector.

We thus calculate with an incoming, positive-energy,
left-helicity tachyonic neutrino.
One obtains the modified sum over spins
${\widetilde \sum}_{\rm spins}$ in the 
matrix element,
\begin{align}
\label{tacTrace}
{\widetilde \sum}_{\rm spins} | \calM |^2 =& \;
\frac{G_F^2}{2} \,
{\rm Tr} \left[
\frac12 \, \left( 1 + \gamma^5 \, \cancel{\tau} \, \cancel{\hat{p}}_3 \right) \;
(\cancel{p}_3 - \gamma^5 \, m_{\nu})  \right.
\nonumber\\[0.1133ex]
& \; \cdot \gamma^5 \, \gamma_\lambda \,
( 1- \gamma^5)
\frac12 \, \left( 1 + \gamma^5 \, \cancel{\tau} \, \cancel{\hat{p}}_1 \right) 
\nonumber\\[0.1133ex]
& \; \left. \cdot 
(\cancel{p}_1  - \gamma^5 \, m_{\nu} ) \, \gamma^5 \, \gamma_\nu  \,
( 1- \gamma^5) \right] \, K^{\lambda\rho} \,.
\end{align}
Here, 
$K^{\lambda\rho} = {\rm Tr} [ ( \cancel{p}_4 + m_e ) \,
\left( c_V \gamma^\lambda - c_A \, \gamma^\lambda \, \gamma^5 \right) 
( \cancel{p}_2 + m_e ) \,
\left( c_V \gamma^\rho - c_A \, \gamma^\rho \, \gamma^5 \right) ]$
is the familiar trace from the outgoing fermion pair.
The decay rate is given by Eq.~\eqref{GammaBasic},
under the replacement
$\frac12 \sum_{\rm spins} | \calM |^2 \to
{\widetilde \sum}_{\rm spins} | \calM |^2$.
The integrals over the momenta of the 
outgoing fermion pair ($\dd^3 p_2$ and $\dd^3 p_4$) are 
done using ($p_2^2 = p_4^2 = m_e^2$)
\begin{align}
\label{resJ}
& J_{\lambda\rho}(q) =
\int \frac{\dd^3 p_2}{2 E_2}
\int \frac{\dd^3 p_4}{ 2 E_4}
\delta^{(4)}( q - p_2 - p_4 ) \,
\left( p_{2 \lambda} \; p_{4 \rho} \right)
\nonumber\\[0.1133ex]
& =
\frac{\pi}{24} \sqrt{1 - \frac{4 m_e^2}{q^2}} 
\left[ g_{\lambda \rho} 
\left( q^2 - 4 m_e^2 \right) 
+ 2 q_\lambda q_\rho
\left( 1 + \frac{2 m_e^2}{q^2} \right) \right] .
\end{align}
It remains to analyze the domain of allowed $\vec p_3$ vectors
[see the ``cupola structure'' in Fig.~\ref{fig2}(b)],
which is defined by the requirement $q^2 > 4 m_e^2$,
for $p_1^\mu = (\sqrt{p_{1z}^2 - m_\nu^2}, 0,0, p_{1z})$.
The dispersion relation 
$E_\nu = \sqrt{\vec p_\nu^{\,2} - m_\nu^2}$ implies that 
\begin{equation}
\label{q2tach}
q^2 = 2 \, \left(
\sqrt{E_1^2 + m_\nu^2} \sqrt{E_3^2 + m_\nu^2} \, \cos \theta
-E_1 E_3 - m_\nu^2 \right) \,.
\end{equation}
Here, $\theta$ is the polar
angle in spherical coordinates,
\begin{equation}
p_3^\mu = (E_3,
|\vec p_3| \, \sin\theta \, \cos\varphi ,
|\vec p_3| \, \sin\theta \, \sin\varphi ,
|\vec p_3| \, \cos\theta) \,.
\end{equation}
Pair production threshold is reached,
for given $E_1$ and $E_3$, by solving 
Eq.~\eqref{q2tach} for $u = \cos\theta$,
setting $q^2 = 4 m_e^2$.
After a somewhat tedious integration over the 
allowed $\vec p_3$ vectors (no masses can be neglected),
one obtains
\begin{subequations}
\label{GammadEdx1}
\begin{align}
\Gamma =& \; \left\{ \begin{array}{cc}
\dfrac{G_F^2 \, m_\nu^6}{128 \, \pi^3 \, m_e^2}
\dfrac{(E_\nu - E_{\thr})^2}{E_{\thr}} &
\qquad E_\nu \gtrapprox E_{\thr} \\[4ex]
\dfrac{G_F^2 \, m_\nu^4}{288 \pi^3} \, E_\nu &
\qquad E_\nu \gg E_{\thr}
\end{array}
\right. 
\end{align}
for the decay rate, and 
\begin{align}
\frac{\dd E_\nu}{\dd x} =& \; \left\{ \begin{array}{cc}
\dfrac{G_F^2 \, m_\nu^5}{64 \, \pi^3 }
\dfrac{(E_\nu - E_{\thr})^2}{E_{\thr}} &
\qquad E_\nu \gtrapprox E_{\thr} \\[4ex]
\dfrac{G_F^2 \, m_\nu^4}{144 \pi^3} \, E^2_1 &
\qquad E_\nu \gg E_{\thr}
\end{array}
\right. 
\end{align}
\end{subequations}
for the energy loss rate.
In the high-energy limit, one may (somewhat trivially) 
rewrite the expressions as follows ($m_\nu = E_1 \, \sqrt{\delta_\nu}$),
\begin{equation}
\label{GammadEdx2}
\Gamma =
\frac{G_F^2 \, E_\nu^5 \, \delta_\nu^2}{288 \, \pi^3} \,,
\quad
\frac{\dd E_\nu}{\dd x} =
\frac{G_F^2 \, E_\nu^6 \, \delta_\nu^2}{144 \, \pi^3} \,,
\quad
E_\nu \gg E_{\thr} \,.
\end{equation}
These results confirm that it is possible to 
use the tachyonic bispinor 
formalism~\cite{ChHaKo1985,Ch2000,%
Ch2002,Ch2002a,JeWu2012epjc,JeWu2013isrn}  for the 
calculation of decay rates of tachyonic particles.

%
% Constraints on the Mass of a Tachyonic Neutrino
%
\section{Constraints on the Mass of a Tachyonic Neutrino}
\label{sec5}

Our threshold relation Eq.~\eqref{mainres}
is based on a Lorentz-covariant dispersion relation.
Only neutrinos with $E_\nu < E_\thr =
\sqrt{2} m_e/\delta_\thr^{1/4}$ survive the possibility of generalized leptonic
\v{C}erenkov radiation over a sufficiently long path length. 
The hypothetical observation of an absence of neutrinos above some
energy $E_\thr$ could thus be interpreted
as a constraint on the neutrino mass.
Let us assume a neutrino mass of $m_\nu = X \, {\rm eV}$,
where $X$ is generally assumed to be of order unity or less. 
Then, threshold is reached for $m_\nu =X {\rm eV}$,
$\delta_\thr =3.67 \times 10^{-24} X^4$,
and  $E_\thr =\frac{522}{X} {\rm GeV}$.

The IceCube
experiment~\cite{AaEtAl2013,AaEtAl2014}  has observed 37 neutrinos having
energies $E_\nu>10 \, {\rm TeV}$ during 3 years of data taking. Three of these events
had energies $E_\nu > 1 \, {\rm PeV}$, and one (often referred to as ``Big Bird'')
had $E_\nu= (2.004 \pm 0.236)\,{\rm PeV}$. 
According to the IceCube collaboration~\cite{AaEtAl2014}, 
the spectrum of the 37 neutrinos is well fitted by a slope
$\sim E_\nu^{-2}$, which includes astrophysical as well as background atmospheric
neutrinos, the latter being exclusively below $0.4\,{\rm PeV}$.  However, their
best fit to the spectrum predicts 3.1 additional events for $E_\nu> 2$ PeV, and yet
none were seen.  Preliminary data for the fourth year includes 17 additional
events, with none seen for $E_\nu > 1 \,{\rm PeV}$~\cite{Bo2015}. 
These facts suggest to the
IceCube authors~\cite{AaEtAl2013,AaEtAl2014} the possibility that there may be
a cutoff for the spectrum for neutrinos above $E\approx 2\,{\rm PeV}$.
The hypothesis is given further support by models which show that the Glashow
resonance~\cite{Gl1960} (resonant ${\overline \nu}_e \, e^{-} \to W^{-} \to
{\rm anything}$) should add between zero and three times
the number of events that appear in the interval $1 \, {\rm PeV} < E_\nu <
2\,{\rm PeV}$ as part of a broad peak centered around 
$6.3 \, {\rm PeV}$~\cite{BaEtAl2014}.   
While evidence for the cutoff is disputed and
alternative explanations have been proposed~\cite{KaTr2015},
the significance of such a cutoff has been analyzed in the
light of superluminal neutrinos~\cite{StSc2014,St2014}.

\begin{figure}[t!]
\begin{center}
\begin{minipage}{0.99\linewidth}
\begin{center}
\includegraphics[width=0.99\linewidth]{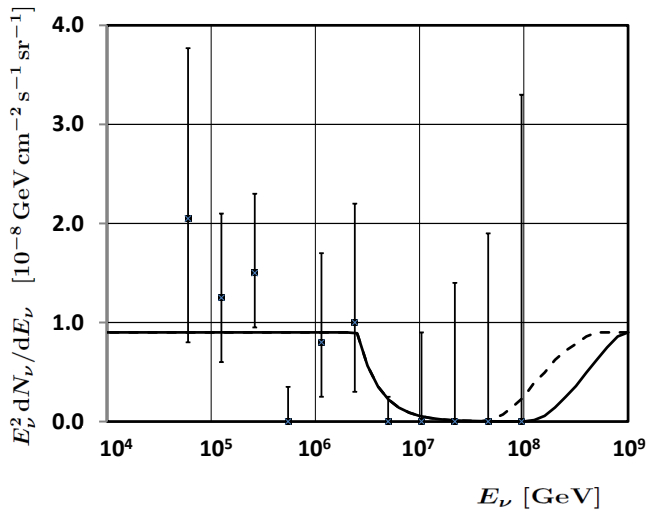}
\caption{\label{fig4} 
Three years of data from the IceCube experiment showing the flux 
$\dd N_\nu/\dd E_\nu$ multiplied by 
$E_\nu^2$ plotted against the neutrino energy $E_\nu$.  The solid and dashed curves
show what would be expected for an $E_\nu^{-2}$ power law for the flux, with an 
$E_\thr=2.5\,{\rm PeV}$
threshold, and two arbitrarily assumed values for the source distance, $L$ (dashed
curve) and $L/2$ (solid curve).  The drop to zero above $E_\thr$ only occurs
for those neutrino flavors having a tachyonic mass consistent 
with a $2.5\,{\rm PeV}$ threshold.}
\end{center}
\end{minipage}
\end{center}
\end{figure}

Let us add a few clarifying remarks here.  First, we note that the plots in the
paper~\cite{AaEtAl2014} refer to the neutrino flux as a function of neutrino
energy; the events were apparently sufficiently well reconstructed so that no
excess neutrino energy in addition to the energy deposited inside the detector
is expected.  Our Fig.~\ref{fig4} is based on Fig.~4 of Ref.~\cite{AaEtAl2014}.
Meanwhile, members of the IceCube collaboration have presented preliminary
evidence for a through-going muon of energy $\geq (2.6 \pm 0.3) \, {\rm PeV}$
which could be interpreted as a decay product of a neutrino of even higher
energy~\cite{EV1,EV2}.  If the through-going muon could indeed be assigned to
an ultra-high-energy neutrino of non-atmospheric origin, then it would push the
conceivable cutoff seen by IceCube to even higher energies, further
constraining the tachyonic mass term of the relevant neutrino flavor.  So far,
the authors of Ref.~\cite{AaEtAl2014} (see the right column on page~4 of
Ref.~\cite{AaEtAl2014}) observe that ``this [the lack of high-energy events]
may indicate, along with the slight excess in lower energy bins, either a
softer spectrum or a cutoff at high energies.''

\color{black}

Assuming $E_\thr \approx 2 \, {\rm PeV}$ 
we would find using Eq.~\eqref{mainres} that 
$\delta_\thr =\big({\sqrt{2} m_e/E_{\thr}}\big)^4\approx 
1.7\times 10^{-38},$ and furthermore,
that $m_\nu= \sqrt{\delta_\thr} E_\thr \approx 0.00026 \, {\rm eV}$ 
(i.e., $-m_\nu^2\approx -6.8\times 10^{-8} \, {\rm eV}^2$) 
for one or more of the three neutrino flavors
(conceivably, the one with the smallest absolute value of $m_\nu^2$). 
A shifted cutoff~\cite{EV1,EV2} of $E_\thr \approx 3 \, {\rm PeV}$,
would be consistent with a 
tachyonic neutrino mass of $m_\nu= 0.00017 \, {\rm eV}$.
One might object that it is not possible to have one (or more) tachyonic flavor
masses ($m^2<0$) and satisfy both neutrino oscillation data and the recent
findings from cosmology for the sum of the flavor masses, i.e., 
$\Sigma m \approx 0.32 \, {\rm eV}$~\cite{BaMo2014,HaHa2013}. However, 
such consistency can be achieved using 3 active-sterile 
$\pm m^2$ (tardyon-tachyon)
neutrino pairs~\cite{Eh2015}.
The curves
in Fig.~\ref{fig4} were generated using an assumed pure $E_\nu^{-2}$ power law for
the flux $N$ beyond the assumed threshold, $E_\thr$.  We then use our
$\dd E_\nu/\dd x$ formula [Eq.~\eqref{GammadEdx1}] for $E_\nu>E_\thr$ to find
the modified $N \, E_\nu^2$ spectrum. 
Good agreement is found with the IceCube data at a threshold
$E_\thr = 2.5 \, {\rm PeV}$, although much more statistics will be needed to
determine if the cutoff is real. 

%
% Conclusions
%
\section{Conclusions}

Three main conclusions of the current investigation can be drawn. 
{\em (i).} As described in Sec.~\ref{sec2},
the assumption of a Lorentz covariant, tachyonic dispersion relation
leads to tight bounds on conceivable
tachyonic neutrino mass terms. The
tachyonic decay rate due to LPCR is most conveniently calculated in the laboratory frame,
because the space-like kinematics involved in the process, which leads to a
non-unique time ordering of the trajectories, 
as discussed in Sec~\ref{sec4}.
{\em (ii).} We may apply the formalism of the tachyonic
bispinor solutions of the tachyonic Dirac 
equation~\cite{ChHaKo1985,Ch2000,Ch2002,Ch2002a} 
recently
developed in Ref.~\cite{JeWu2012epjc,JeWu2013isrn,JeWu2014} to the calculation
of the tachyonic neutrino decay,
as outlined in Sec.~\ref{sec4}.  {\em (iii).} A comparison of
recent IceCube data with the results for the calculated 
tachyonic decay rates reveals
that a tachyonic neutrino could possibly  explain a possible sharp cutoff in
IceCube data, but only if the neutrino flavor involved has a very specific
tachyonic mass. In a more general context,
the calculation of tachyonic thresholds and decay rates 
based on Lorentz-covariant dispersion relations could be 
of phenomenological significance for string theories,
some of which predict the existence of 
tachyons~\cite{Po1998vol1,Po1998vol2}.
The same is true for the precise calculation of the 
tail of the beta decay spectrum, which is 
influenced by a conceivably tachyonic neutrino mass term~\cite{Ci1998}.

\section*{Conflict of Interests}

The authors declare that there is no conflict of interests
regarding the publication of this paper.

\section*{Acknowledgments}

This research has been supported by the
National Science Foundation (Grant PHY--1403973).

\appendix

\end{document}